\documentclass[11pt]{article}
\usepackage{amssymb}
\usepackage{latexsym}
\usepackage{citesort}
\usepackage{url}
\usepackage{a4}
\usepackage{cite}

\newcommand{\mcW}{{\mycal W}}

\newcommand{\gone}{{\widehat{g}}}
\newcommand{\Kone}{{\widehat{K}}}

\newcommand{\nablash}{\nabla{\kern -.75 em
     \raise 1.5 true pt\hbox{{\bf/}}}\kern +.1 em}
\newcommand{\Deltash}{\Delta{\kern -.69 em
     \raise .2 true pt\hbox{{\bf/}}}\kern +.1 em}
\newcommand{\Rslash}{R{\kern -.60 em
     \raise 1.5 true pt\hbox{{\bf/}}}\kern +.1 em}

\newcommand{\pM}{\partial M}

\newcommand{\fg}{{}^4g}

\newcommand{\mcV}{{\mycal V}}

\newcommand{\hyp}{{\mycal S}}
\newcommand{\bhyp}{\,\,\,\overline{\!\!\!\mycal S}}

\newcommand{\mcM}{{\mycal M}}
\newcommand{\mcK}{{\mycal K}}

\newcommand{\bea}{\begin{eqnarray}}

\newcommand{\bel}[1]{\begin{equation}\label{#1}}
\newcommand{\beal}[1]{\begin{eqnarray}\label{#1}}
\newcommand{\beadl}[1]{\begin{deqarr}\label{#1}}
\newcommand{\eeadl}[1]{\arrlabel{#1}\end{deqarr}}
\newcommand{\eeal}[1]{\label{#1}\end{eqnarray}}
\newcommand{\eead}[1]{\end{deqarr}}
\newcommand{\eea}{\end{eqnarray}}

\newcommand{\Ricc}{\mathrm{Ric}\,}

\newcommand{\be}{\begin{equation}}
\newcommand{\ee}{\end{equation}}

\newcommand{\tr}{\mbox{\rm tr}\,}

\newcommand{\eq}[1]{\eqref{#1}}

\DeclareFontFamily{OT1}{rsfs}{}
\DeclareFontShape{OT1}{rsfs}{m}{n}{ <-7> rsfs5 <7-10> rsfs7 <10->
rsfs10}{} \DeclareMathAlphabet{\mycal}{OT1}{rsfs}{m}{n}
\def\scri{{\mycal I}}%
\def\scrip{\scri^{+}}%

\usepackage{amsmath} 

\renewcommand{\theequation}{\thesection.\arabic{equation}}

\let\ssection=\section

\renewcommand{\section}{\setcounter{equation}{0}\ssection}

\newtheorem{defi}{\sc Definition\rm}[section]

\newtheorem{theor}[defi]{\sc Theorem\rm}

\newcommand{\qed}{\hfill $\Box$\bigskip}

\def \Reel{\mathbb{R}}
\def \R {\Reel}

\newcommand{\bM}{\,\overline{\!M}}

\newcounter{mnotecount}[section]

\newcommand{\rmnote}[1]{}

\newcommand{\loc}{{\textrm{loc}}}

\begin{document}
\title
{Existence of non-trivial, vacuum, asymptotically simple
space-times}
\author
{Piotr T. Chru\'sciel\thanks{Supported in part by a grant of the
Polish Research Foundation KBN.}, Erwann Delay\thanks{Supported in
part by the ACI program of the French Ministry of Research.}}
\date{}
\maketitle


\begin{abstract}
We construct non-trivial vacuum space-times with a global
$\scrip$. The construction proceeds  by proving extension results
for initial data sets across compact boundaries, adapting the
gluing arguments of Corvino and Schoen. Another application of the
extension results is existence of initial data which are exactly
Schwarzschild both near infinity and near each of the connected
component of the apparent horizon.
\end{abstract}

\section{Introduction}\label{Sintro}
 In a recent
significant paper~\cite{Corvino} Corvino has presented a  gluing
construction for scalar flat metrics, leading to  the striking
result of existence of non-trivial scalar flat metrics which are
exactly Schwarzschild at large distances; extensions of the
results in~\cite{Corvino} have been announced
in~\cite{CorvinoOberwolfach}. The method consists in gluing
 an asymptotically flat metric $g$ with a
Schwarzschild metric\footnote{Here we mean the metric induced  by
the Schwarzschild metric on the usual $t=0$ hypersurface in
Schwarzschild space-time; we will make such an abuse of
terminology throughout.} on an annulus $B(0,2R_0)\setminus
B(0,R_0)$. One shows that if $R_0$ is large enough, then  the
gluing can be performed so as to preserve the time-symmetric
scalar constraint equation $R(g)=0$, where $R(g)$ is the Ricci
scalar of $g$.

One would like to use the above method to construct vacuum
space-times which admit conformal compactifications at null
infinity with a high degree of differentiability \emph{and} with a
global $\scrip$. Indeed, metrics which are Schwarzschildian, or
Kerrian, near $i^0$ contain hyperboloidal hypersurfaces of the
kind needed in Friedrich's stability theorem~\cite{Friedrich}, and
if the initial data are close enough to those for Minkowski
space-time in an appropriate norm, Friedrich's result  yields the
required asymptotically simple~\cite{FriedrichSchmidt} space-time.
In Corvino's construction there arises, however, an apparent
difficulty related to the fact that if a sequence of data
$(g_i,K_i)$ approach the Minkowski space-time, then the gluing
radius $R_i=R_0(g_i,K_i)$ above could in principle tend to
infinity. This could then lead to hyperboloidal initial data such
that the relevant norm for Friedrich's theorem would \emph{fail}
to approach zero, barring one from achieving the desired
conclusion.

The main object of this letter is to point that the imposition of
a parity condition on the initial data sets considered avoids the
above mentioned problem. We will actually use a slight
variant\footnote{We note that one can very well use the original
method of~\cite{Corvino} to construct the asymptotically simple
space-times, without using our variation of Corvino's method:
Indeed, the arguments given below show that the original argument
of~\cite{Corvino}, in the space of parity-symmetric metrics, will
lead to a gluing radius which can be chosen to be independent of
the metric, for sets of metrics which are bounded in a norm which
controls the decay of the metric and of a finite number of its
derivatives.} of the above construction, which produces extensions
across a boundary $S(0,R)=\partial B(0,R)$, for any \emph{fixed}
radius $R$, for ``small" initial data sets, regardless of whether
or not they are originally arising from an asymptotically flat
initial data set. As a consequence we can produce an infinite
dimensional family of vacuum space-times which are asymptotically
simple in Penrose's sense, with a conformal compactification of
$C^k$ differentiability class for any finite $k$; however, the
method fails to produce $C^\infty$ compactifications. Recall that
the only vacuum, asymptotically simple space-time with a conformal
compactification with a $C^3$ metric known until the examples
presented here was Minkowski space-time\footnote{ The only
non-trivial, with a globally regular $\scri$,
\emph{electro-vacuum} space-times known so far were provided by
the Cutler-Wald metrics~\cite{CutlerWald}.}: this is due to the
fact that the differentiability properties of conformal
compactifications of the space-times constructed by Christodoulou
and Klainerman~\cite{Ch-Kl} or Klainerman and
Nicol\`o~\cite{KlainermanNicolo} are only very poorly controlled
\emph{a priori}.

As another application of the local extension construction we
obtain a family of ``many black holes" initial data sets $(M,K,g)$
with the following property: $M$ is a union of a finite number of
asymptotically flat regions and a compact set $\mcK$. The metric
is \emph{exactly} Schwarzschild or flat on $M\setminus \mcK$.
Further all the Schwarzschild apparent horizons ``${\cal
S}_i:=\{r_i=2m_i\}$" are outside of $\mcK$, so that the metric is
exactly Schwarzschildian in a neighbourhood of each of the ${\cal
S}_i$'s. This provides a family of non-trivial metrics with
``isolated horizons", with the geometry being exactly
Schwarzschildian in a neighbourhood of the horizons.

 Finally, we show that the Corvino-Schoen technique
 can be used
to add \emph{Einstein-Rosen bridge candidates} to manifolds
satisfying $R=0$, assuming in addition  a \emph{local} parity
condition. More precisely, we show that one can deform the metric
to a Schwarzschild one or a flat one in a small neighbourhood of
points satisfying the parity condition. The deformation of the
metric related to the addition of the bridge candidate preserves
the condition $R=0$ and modifies the metric only in an arbitrarily
small neighbourhood of the point at which the bridge is added.
This would allow one to connect pairs of time-symmetric vacuum
initial data without perturbing the metric away from a small
neighbourhood of the bridges, or to create wormholes within a
given initial data set, if one could show that the mass of the
resulting Schwarzschild metrics can be made positive.\footnote{We
suspect that the resulting mass will always be non-positive. This
would make the construction useless for the purpose of adding
bridges. For negative $m$'s one obtains a completely controlled
naked singularity at the initial point. This result might have
some interest in itself.} (Compare~\cite{Joyce,IMP}, where the
conformal method is used for the gluings:  this leads \emph{a
priori} to conformal deformations of the metrics which are small
away from a neighbourhood of the gluing point, but non-zero
\emph{throughout} the manifolds being glued.)  We believe that one
can similarly add \emph{bridge candidates}
to\newcommand{\const}{\textrm{const}} $R=\const\ne 0$ manifolds,
preserving the $R=\const$ condition, but we haven't checked all
the details of such a construction so far.

\section{Extensions of initial data sets}\label{Sstg}
We start by considering the \emph{extension problem}, that is, the
following question: let us be given a vacuum initial data set
$(M,K,g)$, where $\bM=M\cup \pM $ has a compact boundary $\pM$,
with the data $(K,g)$ extending smoothly, or in $C^k(\bM)$, to the
boundary. Does there exist an extension across $\pM$ of $(K,g)$
which satisfies the constraint equations? In the case where $K$
vanishes and $\partial M$ is \emph{mean outer convex} an
affirmative answer can be given by using a method\footnote{Smith
and Weinstein actually assume that $\pM$ is a two-sphere, but this
hypothesis is irrelevant for the discussion here. The special case
when the boundary metric is that of a round two sphere has been
previously considered by Bartnik~\cite{Bartnik93}.} due to Smith
and Weinstein~\cite{SmithWeinstein}. However, it is not clear how
that method can be used to produce extensions which are exactly
Schwarzschildian outside of a compact set. Further, it is even
less clear how to extend this method to accomodate non-trivial
extrinsic curvature.

We wish to point out that the results of Corvino and
Schoen~\cite{Corvino,CorvinoOberwolfach,CorvinoSchoen} can be used
to obtain alternative extension results, \emph{without} the
hypotheses that the boundary is mean outer-convex and  that $K$
vanishes. (It further seems that less differentiability of the
metric is lost with this method, as compared to the
Smith-Weinstein technique; the latter gives $C^k$ extensions of
$C^{2k+1}$ metrics, $k\ge 0$.) Thus, assume we have a solution
$(K,g)\in (C^{k+2}\times C^{k+3})(\bM)$, $k\ge 4$, of the vacuum
constraints on a manifold ${\bM}$ with compact boundary. Let $M_0$
be another manifold such that $\partial M_0$ is diffeomorphic to
$\pM$, and let $M'$ be the manifold obtained by gluing $M$ with
$M_0$ across $\pM$. Let $x$ be any smooth function defined in a
neighbourhood $\mcW$ of $\pM$ on $M'$, with $\pM=\{x=0\}$, with
$dx$ nowhere vanishing on $\pM$, and with $x>0$ on $M_0$. It is
convenient to choose $\mcV:=\mcW\cap M_0$ to be diffeomorphic to
$\pM\times [0,x_0]$, with $x$ being a coordinate along the
$[0,x_0]$ factor.

 Suppose, next, that there exists on
$M_0$ a solution $(K_0,g_0)$ of the vacuum constraint equations
which is in $(C^{k+2}\times C^{k+3})(\bM_0)$; we emphasize that we
do not assume that $(K,g)$ and $(K_0,g_0)$ match across $\partial
M$. Standard techniques allow one to extend $(K,g)$ to a pair
$(\Kone ,\gone )$ defined on  $M_0$ such that
\begin{enumerate}
\item $(\Kone ,\gone )$ remains in $C^{k+2}\times C^{k+3}$;
\item $(\Kone ,\gone )$ coincides with $(K_0,g_0)$ on $M_0\setminus \mcV$;
\item we have
\bel{gder}\|\gone -g_0\|_{C^{k+3}(\mcV)}\leq C
\sum_{i=0}^{k+3}\|\partial_x^i g|_{\partial M}-\partial_x^i
g_0|_{\partial M}\|_{C^{k+3-i}(\partial M)}\;, \ee
\bel{Kder}\|\Kone -K_0\|_{C^{k+2}(\mcV)}\leq C
\sum_{i=0}^{k+2}\|\partial_x^i K|_{\partial M}-\partial_x^i
K_0|_{\partial M}\|_{C^{k+2-i}(\partial M)}\;, \ee
\item
and for all $0\leq i\leq k+1$ it holds that
\begin{eqnarray*}
\lefteqn{
|(\widehat\nabla)^{(i)}\rho(\Kone ,\gone )|_{\gone }+
|(\widehat\nabla)^{(i)}J(\Kone ,\gone )|_{\gone }} && \\
 && \le C\left
(\|\gone -g_0\|_{C^{k+3}(\mcV)}+\|\Kone
-K_0\|_{C^{k+2}(\mcV)}\right)x^{k+1-i}\;.
\end{eqnarray*}
\end{enumerate}
Here $\rho$ is the scalar constraint operator and $J$ is the
vector constraint operator; the constant $C$ might depend upon
$\|\gone -g_0\|_{L^\infty}$ and $\|\Kone -K_0\|_{L^\infty}$. Our
first extension result is obtained under the hypothesis that there
are no $(Y,N)$'s such that $P^*(Y,N)=0$ on $\mcV$, where \be
\label{4} P^*(Y,N)= \left(
\begin{array}{l}
2(\nabla_{(i}Y_{j)}-\nabla^lY_l g_{ij}-K_{ij}N+\tr K\; N g_{ij})\\
 \\
\nabla^lY_l K_{ij}-2K^l{}_{(i}\nabla_{j)}Y_l+
K^q{}_l\nabla_qY^lg_{ij}-\Delta N g_{ij}+\nabla_i\nabla_j N\\
\; +(\nabla^{p}K_{lp}g_{ij}-\nabla_lK_{ij})Y^l-N \Ricc(g)_{ij}
+2NK^l{}_iK_{jl}-2N (\tr \;K) K_{ij}
\end{array}
\right) \;.\ee (Nontrivial fields satisfying $P^*(Y,N)=0$ are
called Killing Initial Data (KIDs)~\cite{ChBeigKIDs}, their
existence implies existence of Killing vectors in the associated
globally hyperbolic vacuum space-time.) If   $(K,g)$ and its
derivatives on $\partial M$ up to appropriate order, as in
\eq{gder}-\eq{Kder}, are sufficiently close to $(K_0,g_0)$ and its
derivatives on $\partial M$,
then the results of~\cite{CorvinoOberwolfach} allow one to
conclude that there exists on $\mcV$ a vacuum initial data set,
close to zero, $(K_0+\delta K,g_0+\delta g) \in \ (C^{k}\times
C^{k})(\overline{\mcV})$, with all derivatives up to order $k$
 coinciding with those of $(K_0,g_0)$ on $\{x_0\}\times\partial M$, and with all derivatives
up to order $k$ coinciding with those of $(K,g)$ on
$\{0\}\times\partial M$.

The above construction has a lot of if's attached, but it does
provide new non-trivial extensions in the following, easy to
achieve, situation: \begin{enumerate} \item $(K,g)$ belongs to a
one-parameter family of solutions $(K_\lambda,g_\lambda)$ of the
vacuum constraint equations on $M$,
\item the vacuum initial data
set $(K_0,g_0)$, assumed above to be defined on $M_0$, is the
restriction to $M_0$ of a vacuum initial data set defined on $M'$,
still denoted by $(K_0,g_0)$, with
\item $(K_\lambda,g_\lambda)$ converging to $(K_0|_{M},g_0|_{M})$
as $\lambda $ tends to zero in $(C^{k+2}\times C^{k+3})(\bM)$.
\end{enumerate}
(Replacing $M$ by a  neighbourhood of $\partial M$, it is of
course sufficient for all the above to hold in a small
neighbourhood of $\partial M$.) In such a set-up, proceeding as
above one obtains an extension for $\lambda$ small enough when
$P^*$ has no kernel on $\mcV$.

The situation is somewhat more complicated when a kernel  is
present, or when working with families of metrics near a metric
which has a kernel, and this is what one has to face when
attempting to construct asymptotically flat metrics with small
mass. We consider a situation where $\bM$ is a smooth compact
submanifold, with smooth boundary, of $M'=\R^3$, with $K_0\equiv
0$, and we let $g_0$ be the Euclidean metric on $\R^3$. The
condition that $\bM$ is a submanifold can be made without loss of
 generality in the following sense: any orientable two dimensional manifold
can be embedded into $\R^3$, and so can a tubular neighbourhood
thereof (this will of course not be an isometric embedding in
general). Replacing $\bM$ by a tubular neighbourhood
$(-x_0,0]\times
\partial M$ of $\partial M$ we can thus identify $\bM$ with a
subset of $\R^3$. We note that the closure of $\bM$ in $\R^3$ will
then have a boundary with two components, $\{-x_0\}\times \partial
M$ and $\{0\}\times \partial M$, but we will ignore
$\{-x_0\}\times \partial M$ if occurring, and consider only
$\{0\}\times
\partial M$, which will be  assumed to be an exterior boundary of $M$ as seen
from infinity.  We assume that $(K,g)$ are close to $(K_0,g_0)$:
\bel{epsineq}\|g-g_0\|_{C^{k+3}(\bM)}+\|K-K_0\|_{C^{k+2}(\bM)}<\epsilon\;;\ee
and that $$g(x)=g(-x)\;, \qquad K(x)=-K(-x)\;;$$ a large family of
such initial data can be constructed by the conformal method. Such
data will be referred to as \emph{parity-covariant}. Clearly the
extensions $(\Kone ,\gone )$ can be constructed as to preserve the
parity properties, and we will only consider {parity-covariant}
extensions.

For definiteness we consider only the case $\Kone =K_0=0$, though
the same argument applies (leading to Kerrian extensions) for
non-necessarily zero but appropriately small $K$'s. The
constructions of~\cite{Corvino} preserve all symmetries of initial
data, so that gluing together ``up to kernel" $\gone $ with
standard (non-translated) Schwarzschild metrics $g_m$ with
$m\in(-\delta,\delta)$, with $\delta \le \min(1, 1/R)$ small
enough, will lead to ``solutions up to kernel" $(\Kone +\delta
K_m=0,\gone +\delta g_m)$ still being parity-covariant. Parity
considerations shows that the center of mass of the resulting
metrics is zero, so that the only obstruction, in the proof
in~\cite{Corvino}, to the requirement that the metric be scalar
flat is the vanishing of the integral of $R(\gone +\delta g_m)$
over $\mcV$. Let $m_0$ denote the naively calculated mass of
$(K,g)$ using the ADM integral over $S(0,R)$, we have $|m_0|\le C
\epsilon$, and by standard identities for the ADM mass integral
one finds
$$\frac 1 {16\pi} \int_{[0,x_0]\times\partial M} R (\gone +\delta g_m)= m-m_0+O(\epsilon^2)\;.$$ If
the reference Schwarzschild metric $g_m$ has mass $m=m_0-\epsilon$
we obtain a strictly negative value of the right-hand-side of the
last equation if $\epsilon $ is small enough. The value
$m=m_0+\epsilon$ leads to a strictly positive value of that
right-hand-side; since the left-hand-side depends continuously
upon $m$, there exists $m\in (m_0-\epsilon,m_0+\epsilon)$ such
that the left-hand-side is zero.

If $K$ does not vanish one needs to choose a constant $0\le
\lambda < 1$ and restrict consideration to initial data sets
satisfying in addition \bel{Kres} |\vec p_0|_\delta \le \lambda
m_0 \;,\ee where $\vec p_0$ is the ADM momentum of $(M,K,g)$. One
then concludes, for $\epsilon$ small enough, using \emph{e.g.\/}
the Brouwer fixed point theorem,  in a manner somewhat similar to
that described in~\cite{Corvino,CorvinoOberwolfach}. Summarizing,
we have proved:

\begin{theor}\label{Textg}
Consider parity-covariant vacuum initial data sets $(K,g)\in
C^{\ell+3}\times C^{\ell+4}$, $\ell\ge 3$, on a compact smooth
submanifold $\bM$ of $\R^3$, suppose that \eq{Kres} holds with
some $0\le \lambda < 1$, and let $\Omega$ be any bounded domain
 containing $\bM$. There exists $\epsilon>0$
such that if \eq{epsineq} holds, then there exists a vacuum
$C^{\ell}\times C^{\ell}$ extension of $(K,g)$ across that part of
$\partial M$ which is homologous to large coordinate spheres in
the asymptotically flat region, with the extension being Kerrian
outside $\Omega$. If $K=0$ then the condition \eq{Kres} is
trivially satisfied by setting $\lambda =0$ (thus no \emph{a
priori} restriction on the sign of $m_0$ is imposed in that case),
and there exists an extension which is Schwarzschildian.
\end{theor}
\qed

\section{Vacuum asymptotically simple space-times}
In this section we wish to show that Theorem \ref{Textg}  can be
used to establish existence of a
 large class of small-data, vacuum space-times with a
global  $\scri$, the conformally rescaled metric being as
differentiable as desired at $\scri$ (though perhaps not
necessarily smooth. Let $M$ in Theorem~\ref{Textg} be a ball
$B(R)$ of fixed radius $R$, we wish to use Friedrich's stability
theorem~\cite{Friedrich} to establish, for data small enough, the
existence of a solution, with a global $\scri$, of the Cauchy
problem with initial data obtained by the extension technique of
Theorem~\ref{Textg}; this proceeds as follows: Let $\hyp$ be any
smooth spacelike hypersurface in Minkowski space-time which
coincides with $\{t=0\}$ for $r\le 2R$, and which coincides with a
hyperboloid $(t-T)^2-r^2=1$, for some $T$, for $r\ge 4R$. Let
$(K,g)$ be any initial data constructed as in the proof of
Theorem~\ref{Textg} by extending from data prescribed on $B(0,R)$,
with $(K,g-g_0)$ small in $(H^{\ell}\times H^\ell)(B(0,2R))$, for
some $\ell \ge 5$, and let $(\mcM,\fg)$ be the maximal globally
hyperbolic development thereof. (We note that the
differentiability condition will hold if we start with initial
data $(K|_{B(0,R)},g|_{B(0,R)})$ which are in $(C^{\ell+1}\times
C^{\ell+2})(B(0,R))$.) If we construct the extension so that the
initial data are Kerrian outside $B(0,2R)$, then the solution will
be a Kerr metric $\fg_Q$ in the domain of dependence of
$\R^3\setminus B(0,2R)$. Here the parameter $Q$ is used to
collectively denote all the global  Poincar\'e charges of the
relevant Kerr metric, that is its ADM four-momentum, angular
momentum, and  center of mass (\emph{cf.,
e.g.}\/~\cite{BOM:poincare,ChAngmom}). We will identify the
hypersurface $\hyp$ with a hypersurface in $\mcM$ as follows:
first, for $r\ge R$ we introduce in Minkowski space-time
Bondi-type coordinates $(u,x,\theta,\varphi)$ by setting $u=t-r$,
$x=1/r$, so that the Minkowski metric $\eta$ becomes \bel{M1}
\eta= x^{-2}\left(-du^2 + 2 dx\, du + d\Omega^2\right)\;,\quad
d\Omega^2 = d\theta^2+\sin\theta\, d\varphi^2\;, \ee with
$\hyp\cap \{x\le 1/(4R)\}$ taking the form \bel{M2}
\hyp=\{u=T+\frac{x}{1+\sqrt{1+x^2}}\}\;.\ee Suppose, first, that
$\fg_Q$ is a Kerr metric in its standard form in retarded
Eddington-Finkelstein coordinates~\cite[p.~879]{MTW}
$(u,r,\theta,\varphi)$, with $|m|+|a|\le \delta$ for a $\delta$
small enough so that the metric has no (coordinate or else)
singularities or horizons throughout the region $r\ge R$, setting
again $x=1/r$ one then has a natural identification between points
on $\hyp$,  understood as a hypersurface in Minkowski space-time,
with a hypersurface in the exterior region of a Kerr space-time,
by using \eq{M2}. In the $(u,x,\theta,\phi)$ coordinate system
$\fg_Q$ takes the form \bel{M3} \fg_Q= x^{-2}\left(-du^2 + 2 dx\,
du + d\Omega^2 + O( x\delta) \right)\;.\ee Further, this form of
$g_Q$ is preserved under small translations, small boosts, as well
as under arbitrary rotations, as long as $|Q|\le \delta$.

If $(K,g)$ are sufficiently close to Minkowski data on $B(R)$ in
$(C^{6}\times C^{7})(\overline{B(0,R)})$, then the initial data
$(K_\hyp,g_\hyp)$ induced by $\fg$ on $\hyp$ will be close to the
Minkowskian data on $\hyp$ in $(H^5_\loc\times H^5_\loc )(\bhyp)$,
where the closure $\bhyp$ of $\hyp$ is taken in the conformally
completed space-time. This implies that the resulting initial data
for Friedrich's conformal equations will be close to those for
Minkowski space-time in the norm needed for Friedrich's stability
theorem~\cite{Friedrich}, yielding global existence whenever
$\epsilon$ in Theorem~\ref{Textg} is made small enough. (We note
that the whole set of initial data needed in Friedrich's theorem
requires divisions by the conformal factor $\Omega$, which
vanishes on $\scrip$, and such operations  could in principle lead
to difficulties when working in Sobolev spaces. However, the data
are Kerrian near Scri, thus conformally smooth there, and the
division by $\Omega$ may be safely performed without the need of
imposing further restrictions on  $\ell$.)

A rough estimate shows that if
$(K|_{\overline{B(0,R)}},g|_{\overline{B(0,R)}})$  are in
$(C^{\ell+5}\times C^{\ell+6})(\overline{B(0,R)})$, $\ell\ge 1$,
then the resulting conformally rescaled metric will be in
$C^\ell(\,\,\,\overline{\!\!\!\mcM})$, where
{}$\,\,\,\overline{\!\!\!\mcM} =\mcM\cup\scrip$ is the conformally
completed space-time. We expect that this can be improved by a
closer inspection of Friedrich's system of equations.

For time-symmetric ($K=0$)  initial data  the existence of a
global $\scri$ follows by covariance of Einstein's equations under
time-reversal. For general data we can repeat the above argument
with the opposite time-orientation  and conclude, decreasing
$\epsilon$ if necessary, that $\mcM$ will possess conformal
completions with a complete $\scrip$ as well as a complete
$\scri^-$.

\section{Initial data with non-connected trapped surfaces (``many
black holes initial data")}

As a second illustration of the extension technique above we
construct time-symmetric initial data for a vacuum space-time with
the following properties:
\begin{enumerate}
\item There exists a compact set $\mcK$ such that $g$ is  a Schwarzchild metric with some
 mass parameter $m$
on each connected component of $M\setminus \mcK$ (in general
different $m$'s for different components);
\item let $\cal S$ denote the usual minimal sphere within the
time-symmetric initial data for the Schwarzschild-Kruskal-Szekeres
manifold, then $M$ contains $2N+1$ such surfaces, with the metric
being Schwarzschild in a neighbourhood of each corresponding $\cal
S$.
\end{enumerate}
In fact, $(M,g)$ will be obtained by gluing together $2N+1$
Schwarzschild metrics with small masses. One can think of $(M,g)$
as an initial data set containing $2N$ black holes. There is,
unfortunately, the usual proviso for such initial data, that it is
not clear whether or not $M$ contains other minimal surfaces than
the Schwarzschild-ones mentioned above; in particular it could be
the case that there is a smooth minimal surface that encloses all
the other ones, so that the outermost apparent horizon will
actually be connected. Further, even if that is not the case, the
intersection of the real event horizon with the initial data
hypersurface $M$ could turn out to be connected, so that the
``many black hole" character of our initial data would actually be
an illusion. We believe that the geometry of $(M,g)$ is striking
enough to make it interesting even if it turned out to describe a
connected black hole after all.

Let us pass now to the description of our construction: choose two
strictly positive radii $0<4R_1<R_2<\infty$, and for $i=1,\ldots,
2N$ let the points
$$\vec x_i \in \Gamma_0(4R_1,R_2):= B(0,R_2)\setminus
\overline{B(0,4R_1)}\;$$ ($B(\vec a, R)$ --- open coordinate ball
centred at $\vec a$ of radius $R$) and the radii $r_i$ be chosen
so that the balls $B(\vec x_i,4r_i)$ are pairwise disjoint, all
included in $\Gamma_0(4R_1,R_2)$. Set \bel{Om1} \Omega:=
\Gamma_0(R_1,R_2) \setminus \left(\cup_i \overline{B(\vec
x_i,r_i)}\right)\;. \ee We shall further assume that $\Omega$ is
invariant under the parity map $\vec x \to - \vec x$. Let
$$\vec M=(m,m_0,m_1,\ldots ,m_{2N})$$
be a set of  numbers satisfying $2m<2R_1$, $2m_0< R_1$, $2m_i <
r_i$, and let the metric $g_{\vec M}$ be constructed as follows:
\begin{enumerate}
\item
on $\Gamma_0(R_1,2R_1)$ the metric $g_{\vec M}$ is the
Schwarzschild metric, centred at $0$, with mass $m_0$;
\item
on $\Gamma_0(3R_1,R_2) \setminus \left(\cup_i \overline{B(\vec
x_i,4r_i)}\right)$ the metric $g_{\vec M}$ is the Schwarzschild
metric, centred at $0$, with mass $m$ ;
\item
on $\Gamma_0(2R_1,3R_1)$ the metric $g_{\vec M}$ interpolates
between the two Schwarzschild metrics already defined above using
a usual cut-off function;
\item on the annuli $\Gamma_{\vec x_i}(r_i,2r_i):=B(\vec
x_i,2r_i)\setminus \overline{B(\vec x_i, r_i)}$ the metric is the
Schwarzschild metric, centred at $r_i$, with mass $m_i$;
\item on the annulus $\Gamma_{\vec x_i}(2r_i,3r_i)$ the metric
interpolates between the two metrics already defined above using a
usual cut-off function;
\item the masses $m_i$, $i=1,\ldots,2N$ are so chosen, and the gluings are so
performed,
  that the resulting metric is symmetric under the parity map
$\vec x \to - \vec x$.
\end{enumerate}
Clearly  $g_{\vec M=0}$ is the flat Euclidean metric on $\Omega$,
in particular it is vacuum.  Obviously the Ricci scalar $R(g_{\vec
M})$ of $g_{\vec M}$ is symmetric under the parity map and
satisfies
$$|R(g_{\vec M})| \le C |\vec M|\;.$$
 Suppose that
\bel{mcondi}|\vec M| \le \delta\;;\ee The results
in~\cite{Corvino} show that for any $k$ there exists $\delta_k $
small enough such that if \eq{mcondi} holds with
$\delta=\delta_k$, then there exists a $C^k$ metric $\hat g_{\vec
M}$ on $\Omega$ which agrees with $g_{\vec M}$ at $\partial
\Omega$, together with derivatives up  to  order $k$, and which is
Ricci scalar flat except for the projection on the kernel of $P^*$
(with $Y=0$, since we are assuming that $K=0$). Parity
considerations as in Section~\ref{Sstg} show that the obstruction
is the non-vanishing of the integral
\begin{eqnarray}\frac 1 {16\pi} \int_\Omega R (\hat g_{\vec
M})& = &m-\sum_{i=0}^{2N} m_i
+O(\delta^2)\;.
\label{Om2}
\end{eqnarray} Fix any set of $m_i$'s,
$i=0,\ldots,2N$, satisfying
$$\sum_{i=0}^{2N}|m_i| \le \delta/4\;.$$ If $\delta$ is small enough
the right-hand-side of \eq{Om2} with $m=\delta/2$ will be strictly
positive; it will be strictly negative with $m=-\delta/2$, by
continuity there exists $m$ such that $\hat g_{\vec M}$ will be
Ricci scalar flat.

The  mass of the solution obtained above, as seen from the end
$r\ge R_2$, might  be very small. One can now make a usual
rescaling  $\hat g_{\vec M}\to \lambda^{-2}\hat g_{\vec M}$,  $r
\to \lambda r$, to obtain any value of the final mass $m$.

We emphasize that the mass parameters $m_i$ and $m_0$ are only
restricted in absolute value, so solutions with some of the
$m_i$'s negative or zero, and/or $m_0$ negative or zero, and $m$
negative, can be constructed. For instance, a zero value of $m_i$
will correspond to metrics which can be $C^k$ matched to a flat
metric on $B(\vec x_i,r_i)$. One can actually also obtain $m=0$:
it suffices to repeat the above argument with prescribed values
$m=0$ and $m_i$, $i=1,\ldots,2N$, adjusting $m_0$ rather than $m$.
This leads to a non-trivial Ricci-scalar flat metric which is flat
on an exterior region $\R^3\setminus B(0,R)$. (Clearly $m=0$
implies that at least one of the $m_i$'s, $i\ge 0$, is negative,
unless they all vanish.)

\section{Adding Einstein-Rosen bridges?}
 In the manifold of the last section one can identify points on pairs of
some of those resulting minimal surfaces which have the same
radius, obtaining --- instead of new asymptotically flat regions
--- wormholes connecting neighbourhoods of the corresponding
points $\vec x_i$. This could be a special case of a more general
construction, which would proceed as follows: Let $p_0\in M$ and
suppose that there exists a coordinate neighbourhood of $p_0 =
\{x^i=0\}$ in which the metric satisfies the parity condition
$g(x)=g(-x)$.
In a manner essentially identical to that described in the
previous section, on a coordinate annulus
$\Gamma_0(\epsilon,4\epsilon)$ we can make a deformation of the
original metric $g$ to a metric which will be a Schwarzschild one,
with small mass $m_\epsilon$, for $r<\epsilon$. We will be
adjusting the parameter $\epsilon$, in order to work on a fixed
set it is convenient to make a rescaling from the annulus
$\Gamma_0(\epsilon,4\epsilon)$ to $\Gamma_0(1,4)$. There exists
$\epsilon_0$ such that for $0<\epsilon<\epsilon_0$ the Corvino
technique produces a deformation which is ``Ricci-scalar flat
modulo kernel". Parity guarantees that the only obstruction to
existence of a solution is the integral
$\int_{\Gamma_0{(1,4})}R\;,$ and, arguing as before, for
$\epsilon$ small enough we can choose the mass $m_\epsilon$ of the
Schwarzschild metric so that the final metric has vanishing scalar
curvature.
In
order to produce an Einstein-Rosen bridge we would have to make
sure that $m_\epsilon$ is positive. We have unsuccessfully tried
to use the harmonic-coordinate calculations of
Bartnik~\cite[Eq.~(5.15)]{Bartnik:mass} to establish positivity.
We do not know whether or not positivity can be achieved, and we
suspect that the resulting mass parameter will always be negative
or zero. One can actually give arguments based on the Penrose
inequality suggesting that such a local construction cannot yield
Einstein-Rosen bridges in general.

\medskip

\noindent{\sc Acknowledgements:} PTC acknowledges the friendly
hospitality of the Albert Einstein Institute, Golm, during part of
work on this paper.

\bibliographystyle{amsplain}
\bibliography{
../../../references/bartnik,%
../../../references/newbiblio,%
../../../references/reffile,%
../../../references/bibl,%
../../../references/Energy,%
../../../references/hip_bib,%
../../../references/netbiblio}
\end{document}